\documentclass[aps,prl,twocolumn,superscriptaddress,showpacs,groupedaddress]{revtex4-1}

\usepackage{amsmath,amssymb,graphicx,color,textcomp}

\newcommand{\ket}[1]{|#1\rangle}

\begin{document}

\title{Quantum Entanglement in Nanocavity Arrays}

\author{T. C. H. Liew}
\author{V. Savona}
\affiliation{Institute of Theoretical Physics, Ecole Polytechnique F\'{e}d\'{e}rale de Lausanne EPFL, CH-1015 Lausanne, Switzerland}

\begin{abstract}
We show theoretically how quantum interference between linearly coupled modes with weak local nonlinearity allows the generation of continuous variable entanglement. By solving the quantum master equation for the density matrix, we show how the entanglement survives realistic levels of pure dephasing. The generation mechanism forms a new paradigm for entanglement generation in arrays of coupled quantum modes.
\end{abstract}

\date{\today}




\maketitle

Entanglement is a key concept in quantum physics and is a crucial resource for quantum information science, particularly within recent schemes based on initial multipartite entangled states~\cite{Raussendorf2001,Menicci2006}. The generation of entangled states of two or more quantum modes typically relies on parametric down-conversion in nonlinear crystals~\cite{Pfister2004,Ferraro2004,Su2007,Yukawa2008} or optical frequency combs~\cite{Pysher2011}. Schemes working at the microscopic scale~\cite{Benson2000,Akopian2006,Mohan2010,Dousse2010} - suitable for integrated devices - are instead always based on the cascaded biexciton-exciton radiative decay in semiconductor nanostructures, and thus restricted to bipartite entanglement.

In systems of weakly nonlinear coupled quantum modes, the interaction energy associated with two quanta is smaller than the broadening introduced by the finite lifetime of the mode. The opposite situation has recently been the object of   theoretical investigation, because of the possibility of engineering strongly correlated many-particle states, and numerous applications ranging from the photon blockade effect~\cite{Imamoglu1997,Birnbaum2005,Verger2006} to the perspective of a quantum simulator~\cite{Hartmann2006,Greentree2006,Angelakis2007,Angelakis2010}. The requirements for a practical realisation of such a strong nonlinearity within a solid state technology are however very stringent, and perhaps the only clear-cut observation of the photon blockade has been reported in a state-of-the-art atomic system~\cite{Birnbaum2005}. We have recently suggested that photons with strongly sub-poissonian statistics can be emitted by a set of coupled modes in the weakly nonlinear regime~\cite{Liew2010}, thanks to the interplay of the weak nonlinearity and quantum interference~\cite{Bamba2011}. We argue that the same mechanism can be more generally applied to the generation of a variety of nonclassical states of many photons - in particular multipartite entangled states.

\begin{figure}[h]
\centering
\includegraphics[width=8.116cm]{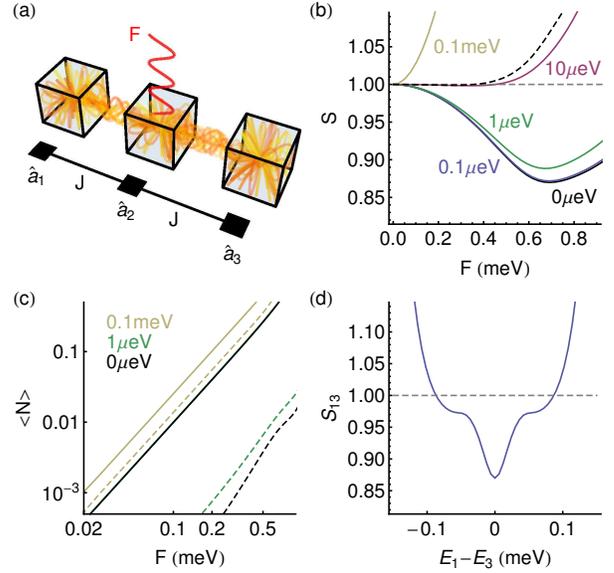}
\caption{(color online) a) Illustration of three coupled confined optically active modes, with optical pumping of the central mode. b) Variation of the entanglement parameter, $S_{13}$, with the pump amplitude for different pure dephasing rates, $\Gamma_P$ (marked on the plot). When the value of $S_{13}$ is less than unity, an entangled state of modes $\hat{a}_1$ and $\hat{a}_3$ is present. The dashed curve shows the value of $S_{12}$ evaluated between modes $\hat{a}_1$ and $\hat{a}_2$. c) Corresponding dependence of the average occupation of modes 1 and 3 (solid curves) and of mode 2 (dashed curves). d) Variation of $S_{13}$ with a non-zero detuning between $E_1$ and $E_3$. We chose $E_1+E_3=-0.06$meV ($E_1=E_3$ in b and c), $E_2=0.08$meV. The slight detuning between the modes $E_{1,3}$ and $E_2$ was found to give the smallest value of $S_{13}$ for fixed $J$ and $\Gamma$ by semi-analytic and numerical optimization (see supplemental material). Note that $J$ is the largest energy scale in the system, such that although the cavities are weakly nonlinear they are strongly coupled.}
\label{fig:S}
\end{figure}

Here, we propose a new paradigm of entanglement generation, which can be implemented in a range of compact solid-state systems including coupled micropillars~\cite{Vasconcellos2011}, coupled mesas~\cite{IdrissiKaitouni2006,Sarchi2008} and coupled photonic crystal cavities~\cite{Yariv1999,Hartmann2006,Greentree2006,Angelakis2007,Gerace2009,Angelakis2010,Ferretti2010}. By accurate theoretical modelling of the open quantum system, we show that continuous variable bipartite entanglement can be generated by an array of three weakly nonlinear spatially confined modes, linearly coupled via quantum tunnelling. The scheme, illustrated in Fig.~\ref{fig:S}a, relies on the quantum interference between distinct excitation pathways influenced by the sensitivity to small nonlinear shifts of the mode energies~\cite{Bamba2011}.
This has the advantage of producing degenerate spatially separated modes, suitable for homodyne detection. We demonstrate entanglement by testing the violation of inequalities~\cite{Duan2000,Simon2000} for separable states, and provide an intuitive interpretation of how the scheme works. Our system is able to reproduce the situation of mode squeezing coupled with beamsplitters~\cite{Braunstein2005} in a compact microscopic system holding promise for an integrated device. The generation mechanism can easily be extended to larger arrays of modes, from which multipartite entanglement is expected.

A general system of three linearly coupled quantum boxes is characterised by: energies $E_n (n=1,2,3)$; photon lifetime $\hbar/\Gamma$, that we assume equal for the three modes; tunneling rate $J$; and nonlinear energy constant $U$. A near-resonant monochromatic pump drives mode 2. We assume that the system lies in the weak nonlinear regime characterised by $U<\Gamma$ and $U<J$~\cite{Imamoglu1997,Birnbaum2005,Verger2006}. Note that excellent control over system geometry, energy detuning and coupling strength has been recently achieved experimentally in the case of semiconductor micropillars~\cite{Vasconcellos2011}. For near-resonant excitation, higher energy modes can be neglected such that each box is described by a single mode. Under these assumptions, the system is described by the Kerr-Hubbard Hamiltonian~\cite{Hartmann2006}:
\begin{align}
\hat{\mathcal{H}}&=\sum_n\left(E_n\hat{a}^\dagger_n\hat{a}_n+U\hat{a}^\dagger_n\hat{a}^\dagger_n\hat{a}_n\hat{a}_n\right)\notag\\
&\hspace{5mm}+J\left(\hat{a}_1^\dagger\hat{a}_2+\hat{a}_2^\dagger\hat{a}_1+\hat{a}_2^\dagger\hat{a}_3+\hat{a}_3^\dagger\hat{a}_2\right)+F\left(\hat{a}^\dagger_2+\hat{a}_2\right)\label{eq:Ham}
\end{align}
where $a_n$ are the Bose annihilation operators of the three modes and $F$ is the optical pump amplitude. This Hamiltonian is written directly in the rotating frame of the pump field, so that $F$ is a constant in time and the energies $E_n$ are expressed relative to the optical pump energy $\hbar\omega_0$. Terms proportional to $U$ describe a Kerr-type nonlinearity. These terms are well suited to model the Kerr nonlinearity induced by the material (that might be enhanced by strong optical confinement) but also a resonant nonlinearity due to exciton-exciton interaction as e.g. in a confined polariton~\cite{Verger2006} system. The quantum optical behaviour of our system is fully described using the master equation for the density matrix, $\boldsymbol{\rho}$:
\begin{align}
i\hbar\frac{d \boldsymbol{\rho}}{dt}&=\left[\hat{\mathcal{H}},\boldsymbol{\rho}\right]+i\frac{\Gamma}{2}\sum_n\left(2\hat{a}_n\boldsymbol{\rho}\hat{a}^\dagger_n-\hat{a}^\dagger_n\hat{a}_n\boldsymbol{\rho}-\boldsymbol{\rho}\hat{a}^\dagger_n\hat{a}_n\right)\notag\\
&\hspace{5mm}+i\frac{\Gamma_P}{2}\sum_n\left(2\hat{n}_n\boldsymbol{\rho}\hat{n}_n-\hat{n}^2_n\boldsymbol{\rho}-\boldsymbol{\rho}\hat{n}^2_n\right),\label{eq:master}
\end{align}
Two Lindblad type terms account for dissipation at a rate $\Gamma$ and pure dephasing at a rate $\Gamma_P$, respectively. The dissipation is caused by the leakage of photons out of the system, while pure dephasing is the result of the coupling to a thermal bath~\cite{Walls1985}. The latter could be due to exciton-phonon scattering in the case of a semiconductor structure. Equation~\ref{eq:master} can be solved numerically for the steady state density matrix using a truncated number state basis~\cite{Verger2006} (see supplemental material for details).

Our aim is to evidence continuous variable entanglement~\cite{Braunstein2005} between the modes in the first and third quantum boxes. In analogy to Bell's result for discrete variable entanglement, continuous variable entanglement is characterised by the violation of an inequality~\cite{Duan2000,Simon2000}:
\begin{equation}
1\leq S_{13}=V\left(\hat{p}_1-\hat{p}_3\right)+V\left(\hat{q}_1+\hat{q}_3\right)\label{eq:S}
\end{equation}
where we have defined the amplitude and phase operators, $\hat{p}_n=\left(\hat{a}_n+\hat{a}^\dagger_n\right)/2$ and $\hat{q}_n=\left(\hat{a}_n-\hat{a}^\dagger_n\right)/(2i)$, respectively. The variance of an operator, $V(\hat{\mathcal{O}})=\langle\hat{\mathcal{O}}^2\rangle-\langle\hat{\mathcal{O}}\rangle^2$, can be extracted theoretically from the density matrix and experimentally measured via homodyne detection.

For our calculations, we use parameters corresponding to exciton-polariton boxes~\cite{Verger2006} although we note that the conclusions of our work also apply to several other physical implementations. It is well-known how to calculate the nonlinear interaction strength~\cite{Verger2006} and we take the value $U=0.012$meV in agreement with experimental measurements~\cite{Kasprzak2007,Amo2009,Ferrier2011}. A range of coupling strengths are possible by varying the separation of the polariton boxes and we choose a coupling strength $J=0.5$meV, which is in agreement with previous theoretical calculations~\cite{Liew2010} and recent experimental measurements~\cite{Vasconcellos2011}. The decay rate $\Gamma=0.044$meV was reported in Ref.~\cite{Wertz2010}.

Figure~\ref{fig:S}b shows the dependence of the parameter $S_{13}$ on the pump amplitude for a range of values of the pure dephasing rate, $\Gamma_P$. For $\Gamma_P=0$, the black curve shows that there is a clear violation of inequality~\ref{eq:S}, corresponding to an entanglement of the modes in the first and third quantum boxes.
In contrast, the modes $\hat{a}_1$ and $\hat{a}_2$ (or symmetrically $\hat{a}_2$ and $\hat{a}_3$) are not entangled, as evidenced by the dashed curve showing the value of $S_{12}$, evaluated from Eq.~\ref{eq:S} by replacing $\hat{a}_3$ with $\hat{a}_2$. While the quantity $S_{13}$ is capable of witnessing entanglement and useful given its experimental accessibility, it is important to note that it does not fulfill the requirements of a direct measure of the amount of entanglement~\cite{Vidal2000}. In fact, there is no unique, universally accepted, measure of the entanglement for our system.

For increasing dephasing rate, the amount of violation decreases and the entanglement is lost at high dephasing rate. Dephasing rates in semiconductor microcavities have been calculated~\cite{Savona1997} and measured~\cite{Houdre2005} in the range of tenths of $\mu$eV. Even for a hypothetical dephasing rate an order of magnitude stronger, we still find that the predicted violation is sufficient for experimental detection.

Figure~\ref{fig:S}c shows the corresponding average populations of the modes in the signal quantum boxes (solid curves) and central box (dashed curves). For small pump amplitudes, corresponding to the linear regime, the populations grow according to a power law as expected. Since $J$ is large, the largest occupations are those of modes $\hat{a}_1$ and $\hat{a}_3$, even though only mode $\hat{a}_2$ is driven. This trend is best understood by expressing the $\hat{a}_n$ operators in terms of eigenmodes of the coupling $J$. Then, similarly to the two-mode system~\cite{Liew2010}, these eigenmodes are driven by the pump in a way that results in destructive interference for the occupation of mode $\hat{a}_2$. Figure~\ref{fig:S}d shows the variation of $S_{13}$ as a function of a finite detuning between the mode energies $E_1$ and $E_3$. The strong resonance at zero detuning is an indication of the underlying quantum interference mechanism. The level of control, required to fabricate a device with such a range of detuning to minimize the entanglement parameter $S_{13}$, is achievable in state-of-the-art arrays of semiconductor micropillars~\cite{Vasconcellos2011}.

We stress that the reported results are also of significance in several other systems. Since the Jaynes-Cummings model can be linked to an effective Kerr nonlinearity~\cite{Boissonneault2009}, Eq.~\ref{eq:Ham} is also applicable to quantum dots embedded in nanocavities and circuit QED systems~\cite{Schoelkopf2008}, where the value of $U$ is related to the cooperativity parameter. In addition, the value of $U$ has been recently evaluated in passive nanocavities~\cite{Ferretti2012}, which represent a particularly promising system given the low decay and dephasing rates. Values of $J$ and $\Gamma_P$ suitable for the present proposal have also been measured for photonic crystal nanostructures. As an example, the coupling of nanocavities has been recently studied in Ref.~\cite{Sato2011} and an upper bound to dephasing rates in quantum dots of $1\mu$eV has been experimentally established~\cite{Langbein2004}.

\begin{figure}[h]
\centering
\includegraphics[width=8.116cm]{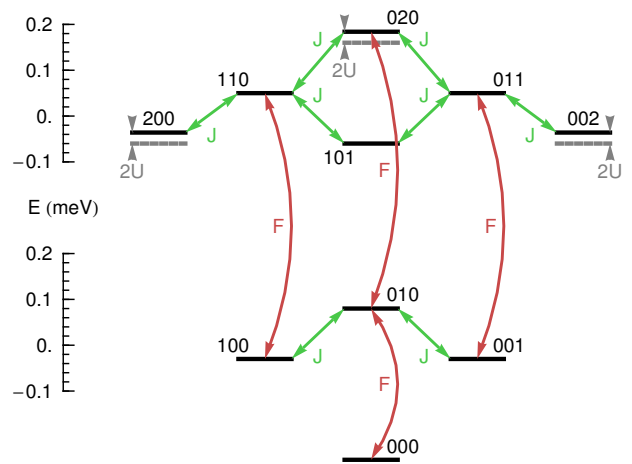}
\caption{(color online) Energy level diagram of the first ten photon number (Fock) states. The states are labelled by the number of photons in each of the three modes. The transitions between modes caused by optical pumping and quantum tunnelling are illustrated by the red (dark gray) and green (light gray) arrows, respectively. The nonlinear shift of states containing two particles can be seen by their difference with the energy levels calculated in the limit $U=0$, which are shown in grey.}
\label{fig:EnergyLevels}
\end{figure}

In order to better understand the origin and the nature of the observed entanglement, we carry out an approximate analysis by expanding the quantum state on a truncated set of photon number states and solving the time-dependent Schr\"odinger equation for this state. This approach does not include the effect of (Lindblad type) dissipation and pure dephasing, and is expected to give an upper bound to the violation of inequality~\ref{eq:S}. The expansion reads:
%
%
\begin{equation}
\ket{\psi}=\sum_{n_1,n_2,n_3}C_{n_1n_2n_3}\ket{n_1n_2n_3}\label{eq:psi}
\end{equation}
where the basis vectors:
\begin{equation}
\ket{n_1n_2n_3}=\hat{a}_1^{\dagger n_1}\hat{a}_2^{\dagger n_2}\hat{a}_3^{\dagger n_3}\ket{000}/\sqrt{n_1!n_2!n_3!}
\end{equation}
represent states with $n_1$, $n_2$ and $n_3$ particles in modes 1, 2 and 3, respectively. For the analysis, expansion~\ref{eq:psi} has to be truncated to a maximum occupation, $N=\sum n_i$. The first ten states, used in expansion~\ref{eq:psi}, are depicted schematically in Fig.~\ref{fig:EnergyLevels}, together with their couplings caused by the pump and tunnelling terms in the Hamiltonian.
The states containing two quanta in the same mode experience slight energy shifts by an amount $2U$ above the bare energy levels (shown in gray) due to the local nonlinear interactions.

\begin{figure}[h]
\centering
\includegraphics[width=8.116cm]{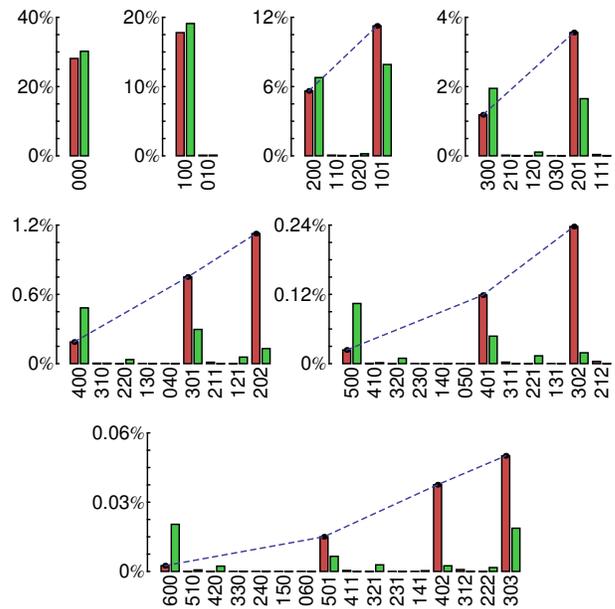}
\caption{(color online) Quantum state of the system expanded onto the particle number states. The values of $|C_{n_1 n_2 n_3}|^2$ are plotted in the linear (red/dark gray bars) and nonlinear (green/light gray bars) regimes. Note that not all the basis states are shown, since due to symmetry $|C_{n_1 n_2 n_3}|^2=|C_{n_3 n_2 n_1}|^2$. The states with $n_2=0$ have the highest occupations and their relative occupations are exactly given by the binomial coefficients in the linear case (dashed lines and black points). The parameters were the same as in Fig.~\ref{fig:S}b with $F=0.8$meV and $\Gamma=0.044$meV.}
\label{fig:Distribution}
\end{figure}

The Schr\"odinger equation, $i\hbar d\ket{\psi}/dt=\hat{\mathcal{H}}\ket{\psi}$, can be solved iteratively under the assumption of small occupations (see the supplemental material for more details) for the steady state (including the effect of particle loss). The coefficients $C_{n_1 n_2 n_3}$ are then calculated and shown in Fig.~\ref{fig:Distribution} for the cases with ($U\neq0$ with green/light gray bars) and without ($U=0$ with red/dark gray bars) nonlinearity.

In accordance with Fig.~\ref{fig:S}c, we observe that the quantum state is in general characterised by very low occupancy of mode 2. Each photon that is initially injected in this mode, tunnels to modes 1 and 3. This behaviour can be easily understood in the linear case ($U=0$), for which the Hamiltonian can be diagonalised exactly. In this case, the Schr\"odinger equation shows that only the mode generated by the operator $\left(\hat{a}_1+\hat{a}_3\right)^\dagger$ is effectively driven by the pump, thus giving rise to a fully separable quantum state, expressed as a linear combination of states $\left(\hat{a}_1+\hat{a}_3\right)^{\dagger N}\ket{000}$ at varying occupancy $N$. Consequently, the relative weights of the coefficients $C_{n_1 n_2 n_3}$, for each given value of the total occupancy $N$, are exactly given by binomial coefficients, as shown by the dashed lines in Fig.~\ref{fig:Distribution}. In the nonlinear regime, the system changes to a state characterised by the green (light gray) bars in Fig.~\ref{fig:Distribution}, where it is clear that states containing particles in both modes 1 and 3 (e.g., $\ket{101}$) are suppressed with respect to the linear case, while those with all particles in the same mode are enhanced (e.g., $\ket{200}$). This result is a consequence of the nonlinear shift when photons occupy the same box, which has an effect on the quantum interference of possible time evolution paths in the Fock basis. As an example, within the manifold of states with $N=2$ occupancy, the couplings of the states $\ket{110}$ and $\ket{011}$ to the states $\ket{200}$ and $\ket{002}$, respectively, change the phase of any time-evolution path passing through those states. If we consider the system initially in the state $\ket{011}$ for example, then to reach the state $\ket{101}$ two possible options are clear from Fig.~\ref{fig:EnergyLevels}: the direct path $\ket{011}\mapsto\ket{101}$ or the path $\ket{011}\mapsto\ket{020}\mapsto\ket{110}\mapsto\ket{101}$. The relative quantum phase of the two paths is affected by coupling to the state $\ket{200}$, which in presence of the small nonlinear shift of this state induces a destructive interference of the two paths and suppression of the state $\ket{101}$. We are left with a situation where the detection of at least one photon in either signal mode, 1 or 3, grants that no photon will be detected in the other mode. This result solely depends on the nonlinearity in modes 1 and 3. We have verified that the parameter $S_{13}$ experiences negligible change when the nonlinearity in mode 2 is removed.

In summary, arrays of coupled photonic modes are able to display striking quantum correlations despite their modest nonlinearity in the low occupation limit. This allows continuous variable entanglement to be generated between degenerate spatially separated modes that are coupled via quantum tunnelling, in a way that is robust to typical decoherence rates in these systems. The set of three coupled modes here described serves as a building block that can be repeated on an array of modes with appropriate topology, which could be further controlled using electric or magnetic fields~\cite{Zhang2009}. This sets a viable paradigm for the generation of multiparty entanglement in arrays of quantum boxes on a single device.

Our work was supported by NCCR Quantum Photonics (NCCR QP), research instrument of the Swiss National Science Foundation (SNSF).

\end{document}